\documentclass[conference,compsoc]{IEEEtran}
\usepackage[frozencache,cachedir=_minted-main]{minted}
\usepackage{pifont}
 \usepackage{amsmath}
\usepackage{comment}
\usepackage[ruled,linesnumbered]{algorithm2e}
\usepackage{hyperref}
\usepackage{cleveref}
\usepackage{tikz}
\usetikzlibrary{arrows.meta}

\ifCLASSOPTIONcompsoc
  \usepackage[nocompress]{cite}
\else
  \usepackage{cite}
\fi

\usepackage{color}
\usepackage{amssymb}
\usepackage{multirow}
\usepackage{booktabs}

\usepackage{graphicx}
\usepackage{subcaption}

\usepackage{listings}
\usepackage{xcolor}
\setminted[c]{
frame=lines,
framesep=2mm,
baselinestretch=0.8,
fontsize=\footnotesize,
tabsize=2,
linenos,
breaklines}

\definecolor{todocolor}{rgb}{0.9,0.1,0.1}
\definecolor{dxcolor}{rgb}{0.7,0.7,0.3}
\definecolor{bxlcolor}{rgb}{1.0,0.0,0.5}
\definecolor{lxcolor}{rgb}{0.7,0.3,0.7}
\definecolor{codegreen}{rgb}{0,0.6,0}
\definecolor{codegray}{rgb}{0.5,0.5,0.5}
\definecolor{codepurple}{rgb}{0.58,0,0.82}
\definecolor{backcolour}{rgb}{0.95,0.95,0.92}
\definecolor{emphcolor}{rgb}{0.58,0,0.29} 
\definecolor{highlight}{rgb}{0,0,1}
\definecolor{highlight2}{rgb}{1,0.64,0}

\definecolor{packagecolor}{rgb}{0.5, 0.0, 0.5}  
\definecolor{descriptioncolor}{rgb}{0.0, 0.5, 0.5} 
\definecolor{bannedcolor}{rgb}{0.85, 0.1, 0.1} 

\usepackage{tcolorbox}

\begin{document}

\title{Compatibility at a Cost: Systematic Discovery and Exploitation of MCP Clause-Compliance Vulnerabilities}

\author{
\IEEEauthorblockN{Nanzi Yang, Weiheng Bai, Kangjie Lu}
\IEEEauthorblockA{
University of Minnesota \\
Minneapolis, USA \\
yang9467@umn.edu, bai00093@umn.edu, kjlu@umn.edu
}
}

\maketitle

\begin{abstract}
The Model Context Protocol (MCP) is a recently proposed interoperability standard that unifies how AI agents connect with external tools and data sources. By defining a set of common client-server message exchange clauses, MCP replaces fragmented integrations with a standardized, plug-and-play framework. However, to be compatible with diverse AI agents, the MCP specification relaxes many behavioral constraints into optional clauses, leading to misuse-prone SDK implementation. We identify it as a new attack surface that allows adversaries to achieve multiple attacks (e.g, silent prompt injection, DoS, etc.), named as \emph{compatibility-abusing attacks}.

In this work, we present the first systematic framework for analyzing this new attack surface across multi-language MCP SDKs.
First, we construct a universal and language-agnostic intermediate representation (IR) generator that normalizes  SDKs of different languages. Next, based on the new IR, we propose auditable static analysis with LLM-guided semantic reasoning for cross-language/clause compliance analysis. Third, by formalizing the attack semantics of the MCP clauses, we build three attack modalities and develop a modality-guided pipeline to uncover exploitable non-compliance issues.


Our approach shows practical impact: across ten SDKs, it detects 1,265 potential risks. Given the large volume of detected risks, we randomly sampled and submitted 26 reports, of which 20 have been acknowledged and 5 triaged as high-priority. MCP maintainers acknowledged the high acceptance rate and the impracticality of reporting the overwhelming issues individually, as such they invited our tools for integration in the MCP conformance-testing Specification Enhancement Proposal (SEP).  These results confirm that \emph{compatibility-abuse} attacks are  practical and can be pervasive; it is not simply programming errors but rooted in the fundamental tension between agent diversity and MCP’s standardization requirements.

\end{abstract}



\IEEEpeerreviewmaketitle


\section{Introduction}

As a \emph{USB-C port} for AI applications, the Model Context Protocol (MCP) provides a standardized interface for connecting AI models with external tools and data sources~\cite{mcp_usb}. Backed by official language-specific SDKs that implement the MCP standard and hide integration details, MCP replaces fragmented model-specific integrations with a single universal protocol~\cite{anthropic_mcp_introduction}. Major AI providers and developer platforms have quickly embraced MCP, such as OpenAI, Anthropic, and even GitHub’s Copilot, leading to thousands of MCP servers and rapid ecosystem growth~\cite{openai_support_mcp, anthropic_mcp_introduction, copilot_mcp}. Consequently, MCP is becoming a cornerstone of modern AI agent ecosystems.

In its design, MCP follows a client–server architecture~\cite{mcp_architecture}. An MCP client (embedded in the AI application) maintains connections to one or more MCP servers, discovers the services those servers expose (e.g. tools, data resources, prompt templates), and relays requests on behalf of the backend Large Language Model (LLM). Each MCP server implements the protocol to expose a set of external capabilities and responds to client requests. At its core, the MCP specification defines a set of \textit{clauses} governing the \textbf{message exchange} between client and server, essentially specifying \textbf{who} (client or server) is allowed to send \textbf{what} message at \textbf{when}. By standardizing this communication contract, MCP cleanly decouples the LLM from external services, allowing any compliant client to work with any compliant server (regardless of programming language or platform) in a plug-and-play fashion.

However, the \textbf{standardization vs. diversity} of AI agents presents a delicate tension in MCP's design.
Typically, AI agents are broadly diverse; for instance, they interact with heterogeneous servers, tools, and resources via different methods across scenarios. However, MCP aims to unify these interactions through a common protocol.
To remain broadly compatible across varied use cases, the MCP specification adopts an RFC-style approach with only a minimal core of unconditional requirements and many optional or conditional clauses. According to the RFC keyword explanation, should is a recommended keyword but not an absolute requirement~\cite{rfc_keyword}. It may exist valid reasons in particular circumstances to ignore the specific item~\cite{rfc_keyword}. 
Consequently, our analysis of the specification shows that only about \textbf{21.5\%} of MCP's clauses are absolute \textbf{MUST} requirements, whereas the other \textbf{78.5\%} are marked as non-mandatory (conditional \textit{MUST}, \textit{SHOULD}, \textit{MAY}, etc.). This quantifies MCP's compatibility-prioritized design rationale, which is to support diver agent scenarios. The majority of the protocol's behaviors are \textit{optional} rather than strictly enforced. 


This high compatibility design introduces a critical gap when it comes to SDK implementations. 
From the protocol designer’s perspective, MCP’s high compatibility (many optional clauses) is a deliberate choice to accommodate agent diversity. 
But from an SDK developer’s perspective, optional clauses are effectively features that can be omitted.
Because these clauses are not strictly required, SDK developers often use their discretion to implement or skip them.
For example, the MCP specification states: \emph{``When the list of available tools changes, servers that declared the \textsc{listChanged} capability SHOULD send a notification''}. As should is a recommended keyword but not an absolute requirement, it may exist valid reasons in particular circumstances to ignore the specific item~\cite{rfc_keyword}. Besides, according to the MCP official document~\cite{mcp_schmea}, it is an optional notification from the server to the client. We found that the TypeScript SDK does implement sending this notification, whereas the Python MCP SDK does not. In general, each language-specific SDK may implement only a subset of the MCP clauses, especially the non-mandatory ones, according to the developers’ judgment.

This inconsistent enforcement of high-compatibility clauses opens a new attack surface. 
From an attacker’s perspective, each MCP clause can be viewed as a security constraint on the system’s behavior—thus, a missing or unenforced constraint becomes an exploitable opportunity. 
For instance, because the Python SDK omits the above \textsc{listChanged} notification clause, a malicious MCP server could silently alter its tool descriptions; the client would automatically accept the changed description and inject it into the LLM’s context without any alert or oversight. 
In other words, by abusing an unimplemented clause, the attacker achieves a silent prompt injection. We refer to this class of exploits as \emph{compatibility-abuse attacks}.
Importantly, this attack surface is not an incidental bug; however, it stems directly from MCP’s compatibility-first design philosophy. 
By prioritizing broad adoption with many optional protocol rules, the specification inherently leaves enforcement gaps that attackers can exploit. 
In essence, the agent diversity that MCP accommodates is the very reason these vulnerabilities exist, making compatibility-abuse attacks intrinsic to MCP’s design.




Even after recognizing these \emph{constraint gaps} and their root cause, systematically identifying all such issues is non-trivial. 
Performing a thorough analysis requires adopting multiple viewpoints, including the protocol designer, SDK developer, and attacker, and linking them together. 
In our work, we set out to automate this analysis and face three main technical challenges:



First, although SDKs of different languages implement the same MCP specification, the specification constrain only their external protocol behaviors, not how those behaviors are implemented in each language. In practice, SDKs realize the MCP semantics using very different code structures and abstractions. Furthermore, the number of SDK is increasing,  without a common representation, every new SDK would require its own bespoke analysis logic, making the tool not scalable as the SDK grows. As a result, the first challenge is to normalize SDK implementations into a canonical, language-agnostic intermediate representation (IR) of their MCP-relavant behavior, such that compliance checks can be reused across all SDKs.

Second, even with the universal IR, checking non-implemented clauses is still challenging. Implementation gaps can occur for any clause in any SDK, forming a broad matrix of cases to check. 
Checking non-implementation across this clause–SDK matrix must account for language-specific coding styles and remain auditable. Two mainstream approaches exist for analyzing these non-implementations: pattern-based static analysis and LLM-based analysis. A pure pattern-based static analysis would risk a pattern explosion, such as too many code patterns to match across different languages and clauses, while an LLM-based analysis could reason about code semantics but might hallucinate or lack transparency. As a result, we need to propose a hybrid approach that combines their unique advantages while avoiding their respective drawbacks.




Third, not every missing clause implementation constitutes an exploitable vulnerability. Many omissions may have no security impact and simply reflect benign differences in SDK behavior. 
A key challenge is to discern which spec compliance gaps actually open the door to attacks. 
In fact, existing security scanners and agent testing tools tend to rely on hard-coded attack templates or heuristics~\cite{mcp-scan,mcp-scan-web,prisma-airs,datedome-mcp,prompt-security}, which cannot adequately distinguish mere compliance issues from true vulnerabilities. These template-based methods are inherently limited; they carry a bias toward known attack patterns and struggle to anticipate novel exploits arising from the unique semantics of MCP clauses. We need a principled way to filter out non-security-relevant compliance violations and home in on the dangerous ones. In other words, the analysis must understand the semantics of each clause and how its absence could be leveraged by an attacker, rather than simply flagging all deviations from the spec.

To address these challenges, we provide a systematic analysis framework with three key techniques. To solve the first challenge, we propose \emph{universal IR generator}. Our key observation is that typical MCP clauses essentially specify a \emph{conditional action}. For example, in the \texttt{ToolChangeNotification} clause, the condition is available tool changes, and the action is to send a specific notification. At the program level, this corresponds to a function call guarded by a specific condition. This abstraction is cross-language. We construct a language-agnostic IR by extracting a conditional-call graph from each SDK: a catalog of functions (potential MCP actions) and the conditions under which they are invoked. This normalized representation allows us to apply the same non-implemented checks across different SDKs.



To solve the second challenge, we propose \emph{hybrid static-LLM analysis}. Our key observation is that static analysis is good at code slicing, while LLM is good at semantic understanding and reasoning. Therefore, we combine these via a hybrid design: static analysis first slices code via IR, shrinking the search space from the whole codebase to a finite subgraph; Within them, LLM performs semantic reasoning about the invoked function's intent (action) and guard satisfaction (condition). This hybrid approach makes LLM reasoning under the guidance of static analysis, which avoids the hallucination of LLM reasoning on entire codebases or writing hundreds of brittle patterns.


To solve the third challenge, we propose \emph{modality-based exploitability analysis}. Our key observation is that MCP’s clauses constrain two dimensions of agent behavior: (1) \emph{payload}: the content of messages (what data is sent) and (2) \emph{timing}: the timing or sequence of messages (when or if something is sent). An omitted clause becomes exploitable if an attacker gains control over either payload or timing as a result. If not, we treat it as a normal bug. This method filters the non-implementations down to those with potential security impact.

For non-implementation analysis, we adopt our tool to all 10 MCP official SDKs, with 1,270 non-implementations identified. Based on this, our exploitation analysis identified 1,265 potential risks. We measure the accuracy of the non-implementation analysis result, which shows an average 14\% false positive rate, 13.5\% FN rate, 86\% precision, and 87.0\% recall. These indicate that our analysis achieves acceptable accuracy.

For the exploitability analysis, we adopt our tools to each non-implementations, 1,265 of all 1,270 non-implementations are marked as exploitable. Given the large volume of potential risks, we randomly sampled and submitted 26 reports through the proper channel.  As of date, we disclosed 3 issues to the MCP community via the official security channel~\cite{mcp_security} with all confirmed. For SDKs, we disclosed 20 issues to the Python SDK; 15 were acknowledged with a ``\emph{ready for work}'' label by maintainers, and 5 were triaged as high priority (3 P0 and 2 P1). Beyond Python, maintainers of the Go and TypeScript SDKs each confirmed one of our reports. 

Considering the high accept/report rate and the large number of non-submitted potential risks, community maintainers invited our tool to integrate with the MCP community beyond individual SDKs. Community maintainers show that many of our findings align with an ongoing standards enhancement proposal (SEP). After showing the intention of open-sourcing our tool, the maintainers want to integrate our tool as one part of the conference-testing SEP. 

Beyond individual SDKs, our tool has been invited to contribute to the broader MCP ecosystem. Community maintainers show that many of our findings align with an ongoing standards enhancement proposal (SEP). After showing the intention of open-sourcing our tool, the maintainers want to vote and integrate our tool as one part of that SEP. These engagements show that our tools have a promising community impact.

Overall, this paper makes the following contributions:


%
\noindent $\bullet$ \textbf{Novel Attack Surface}: We show why the agent's diversity feature forces the MCP to achieve high compatibility, and how attackers can abuse these high-compatibility clauses to launch multiple attacks in real SDKs, which we term \emph{compatibility-abuse attacks}. This attack surface is intrinsic to the MCP design.

\noindent $\bullet$ \textbf{Systematic Analysis}: We propose a language-agnostic, LLM-static hybrid, modality-based three-step analysis framework. It normalizes different SDKs into universal IR, making the auditable, scaling compliance checks across clauses $\times$ SDKs, and assessing exploitability based on abstract attacking modalities. 


\noindent $\bullet$ \textbf{Real Community Impact}: Our tools discovered more than a thousand issues. Our findings have been acknowledged and triggered extensive discussions in the community. Our tool is being integrated into the community's ongoing conformance-testing SEP. 


\section{Background}
\label{background}

\subsection{AI Agents}
\label{agents}



An AI Agent is a large language model (LLM), configured with instructions and tools~\cite{openai_agents_sdk_python}. In practice, agents are deployed for highly different tasks (e.g., writing, research \& analysis, programming, etc.)~\cite{openai_gpts_platform}. To achieve this, agent implementations vary in their prompts and the external services they rely on (e.g., resources, tools, etc.). In addition, agents are deployed on vendor-specific LLM platforms with distinct APIs (e.g., OpenAI GPTs, Claude Agents, Gemini Gems)~\cite{openai_gpts_platform, claude_agents,gemini_gems}. Taken together, the diversity across tasks, implementations, and vendors is an intrinsic feature of the agent paradigm.

These diversity features pose fundamental challenges for the design of any protocol that aims to unify agent interactions. On the one hand, such a protocol must provide common standards to enable interoperability across heterogeneous agents deployed on multiple LLMs. On the other hand, it cannot over-constrain agent behavior, since the very diversity of tasks, tools, and platforms demands high-compatibility in how the specification is implemented. 

\begin{figure}[t]
	\centering
	\includegraphics[scale=0.25]{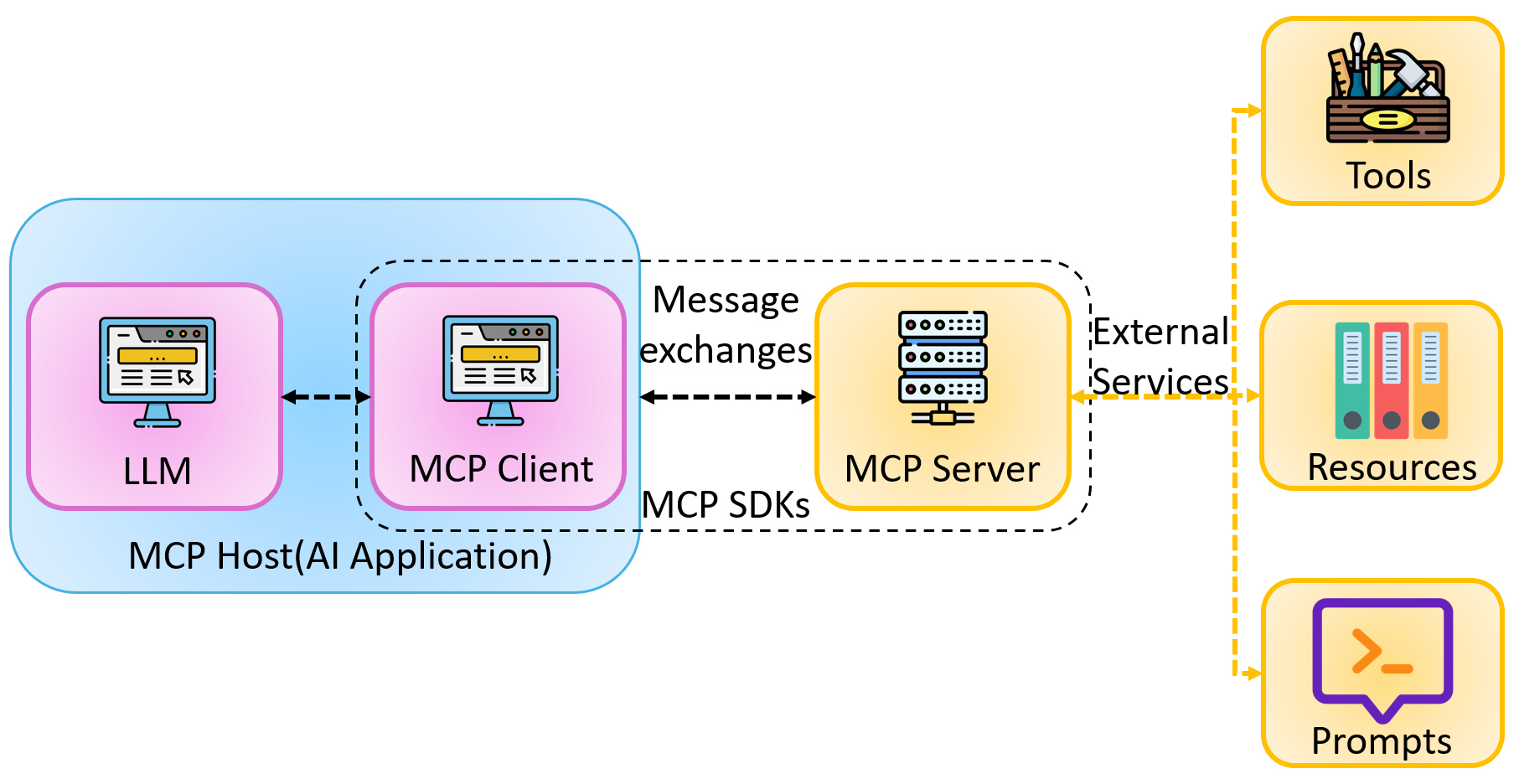}
    \caption{MCP Architecture}
	\label{fig:mcp_architecture}
\end{figure}

\subsection{MCP Architecture}

To unify how AI agents interact with external services across vendor platforms, Anthropic proposed the Model Context Protocol (MCP), which has since been widely adopted by different vendors~\cite{mcp_clients}. As shown in~\cref{fig:mcp_architecture}, MCP groups the LLM together with an MCP client into an MCP host. The MCP client acts as an application-side adapter: it sends requests from the LLM to external services and forwards their responses back to the LLM. On the other side, an MCP server exposes a set of external services (e.g., tools, resources, and prompt templates) and processes MCP requests from the host, returning structured responses. As a result, MCP acts as a shim layer between LLMs and external services: it standardizes and specify the format and semantics of the messages exchanged between them, eliminating the need to maintain model-specific message exchange mechanisms.

In its design, MCP standardizes and specifies the message exchanges between clients and servers through a set of RFC-style clauses. Each clause is written using RFC 2119/8174 keywords (e.g., \texttt{MUST}, \texttt{SHOULD}, \texttt{MAY})~\cite{mcp_specification}. For example, MCP states: “\emph{When the list of available tools changes, servers that declared the LISTCHANGED capability SHOULD send a notification}.” Here, \texttt{SHOULD} is an RFC keyword indicating a  recommendation: a MCP server is expected to send a notification message whenever its set of tools changes. To help developers build MCP clients and servers that follow these clauses, MCP provides a set of official SDKs in different languages, each encode the same set of rules by different developers.

To embrace the diversity of agent adopting scenario, MCP must accommodate agents whose message exchange requirements can be actually conflicting. For example, agent A want to change its tool set dynamically and require notifications to inform the clients tools updating. However, agent B just use a static tool set and don't need the notification mechanism at all; if the specification marked all tool-related messages as unconditional MUST, it would force Agent B to implement irrelevant semantics, creating a direct conflict. As a result, MCP designer use the SHOULD keyword, which mean that there may exist valid reasons to ignore a particular item~\cite{rfc_keyword} As described by MCP official document, this notification is \emph{``an optional notification from the server to the client''}~\cite{mcp_schmea}.

From an SDK developer's perspective, ``optional'' semantics mean that these mechanisms may or may not be implemented. For example, the Python MCP SDK does not implement the notification-sending mechanism for when adding or changing tools. If a malicious MCP server silently change tools with embedded malicious instructions, the MCP client receives no LISTCHANGED notification and therefore cannot detect that new tools have appeared. The LLM can then be injected by these tools without any explicit signal to the user or application, creating a silent prompt-injection attack surface (see~\cref{motivation} for details).

\section{Motivation}
\label{motivation}



Diversity is a unique feature intrinsic to the AI agents. As a protocol that aims to unify and satisfy all agent interactions across different requirements, MCP adopts a compatibility-first design: As the guardrail for message exchanges, it keeps only a small core of mandatory MUST clauses, and relies on a large number of optional clauses.  However, in implementation, these optional clauses are encoded unevenly across different SDKs. From the attacker's perspective, missing clauses mean omission of guardrails, which can be exploited to launch attacks. In the following section, we first quantify the number of optional/mandatory clauses in the specification itself, then provide a real example to show how non-implemented clauses can be exploited to launch attacks.





\subsection{Quantifying MCP Clauses}

In the MCP official document, it states: “\emph{Whether you’re building an AI-powered IDE, enhancing a chat interface, or creating custom AI workflows, MCP provides a standardized way to connect LLMs with the context they need.}”~\cite{mcp_specification}. This description highlights that MCP is explicitly designed to support diverse agent use cases through a single standardized interface. In our analysis, this compatibility-first philosophy manifests in how the specification distributes requirement levels, as we quantify below.



\begin{table}[t]
\centering
\caption{Quantification of MCP clauses}
\label{tab:mcp-compat-firstcol}
\scriptsize
\begin{tabular}{cccc}
\toprule
\textbf{Compatibility} & \textbf{RFC Keywords} & \textbf{Count} & \textbf{Share} \\
\midrule
Mandatory & Unconditional MUST & 59 & 21.5\% \\ 
\multirow{2}{*}{Optional} & Conditional MUST & 78 & 28.4\% \\ & Non-MUST & 138 & 50.2\% \\  \bottomrule
\end{tabular}
\end{table}

In the latest stable MCP protocol specification (2025-06-08 version)~\cite{mcp-versions}, there are 275 clauses in total. We perform a quantitative analysis of their normative strength. Following the RFC keyword semantics, we treat all non-\texttt{MUST} clauses (e.g., \texttt{SHOULD}, \texttt{MAY}) as optional, since they explicitly allow non-implementation under specific circumstances~\cite{rfc_keyword}. 

For \texttt{MUST} clauses, we further distinguish two cases. Unscoped \texttt{MUST} clauses, which apply unconditionally, are classified as mandatory. In contrast, scoped or conditional \texttt{MUST} clauses (e.g., those that only apply when a server satisfies a particular requirements) are classified together with optional clauses, because an implementation can legitimately avoid these requirements by not adopting the corresponding scenario. For example, the MCP states: \emph{``When using the HTTP transport, servers MUST implement OAuth 2.0''}. This applies only to the agent whose server with HTTP transport, and does not apply to an agent using local stdio transport. These clauses can be treated as somehow optional as they couldn't adopting to all scenarios.

Using this scheme, we label all MCP clauses as either mandatory or optional for our subsequent measurements. The results are summarized in~\cref{tab:mcp-compat-firstcol}. In total 275 clauses, 138 of them are no MUST item, and 137 are MUST item. In other words, if we treated all MUST as mandatory, 49.8\% clauses are optional. Furthermore, in the 137 MUST item, only 59 of them have no adopting scope, and 78 of them are limited with specific scope. Consequently, only 21.5\% MCP clauses are non-optional, and 78.5\% of them can be adopted according to agent requirements. From these quantitive analysis, we demonstrate that MCP do use the compatibility-first design, which is driven by the diversity requirements of AI agents.



\subsection{Motivating Example: Silence Injection via Missing Notifications}

Given MCP’s prevalence of optional clauses, SDK developers are not required to implement all of them and may legitimately leave many unimplemented in specific SDKs. From an attacker's perspective, each clause can be semantically regarded as a guardrail that specifies which messages should be exchanged and under what conditions. When a clause is not implemented, the corresponding guardrail is absent, and the related messages are no longer protected. Attackers can exploit these missing guardrails to mount various attacks. We refer to such exploits as \emph{compatibility abuse attacks}.


\noindent\textbf{Threat model.}
In this section, we assume a malicious or compromised MCP server. The attacker controls the server side and leverages non-implemented MCP clauses to attack the MCP client and the LLM behind the client. We assume that the MCP client and the backend LLM are under the control of the victim user.

This assumption is proposed for two reasons. First, there already exist many public third-party MCP servers provided by community repositories~\cite{mcp_repository,awesome-mcp-servers,official-mcp-servers}. These servers are developed by different developers and vary in quality. They can not be uniformly assumed to be benign. In contrast, MCP clients are typically provided by LLM vendors as a part of the LLM platform official integrations~\cite{mcp_clients}. Thus, from a typical user’s perspective, it is natural to trust the client/LLM stack while connecting to potentially untrusted MCP servers.

Second, existing MCP vulnerability detectors adopt the same assumptions; they all target on detecting and defending against the malicious MCP server threats~\cite{mcp-scan,mcp-scan-web, prisma-airs,datedome-mcp,prompt-security}. Based on these two considerations, we adopt a malicious-server assumption in our threat model.

\begin{figure}[t]
\small
\centering
\begin{subfigure}{0.45\textwidth}
\begin{minted}[fontsize=\scriptsize,highlightlines={1,4},breaklines=true, samepage]{c}
async def send_tool_list_changed(self) -> None:
    """Send a tool list changed notification."""
    await self.send_notification(types.ServerNotification
        (types.ToolListChangedNotification()))
\end{minted}
\caption{The definition of send\_tool\_list\_changed method.}
\label{fig:send-tool-list-changed}
\end{subfigure}

\vspace{10pt}
\begin{subfigure}{0.45\textwidth}
\begin{minted}[fontsize=\scriptsize, breaklines=true, highlightlines={5,9,10}, samepage]{c}
    def add_tool(
    ...
    ) -> Tool:
        """Add a tool to the server."""
        tool = Tool.from_function(
        ...
        )
        ...
        self._tools[tool.name] = tool
        return tool
\end{minted}
\caption{The definition of add\_tool method.}
\label{fig:define-add-tool}
\end{subfigure}
\caption{The hook omission in Python SDK.}
\label{fig:hook-omission}
\end{figure}

\noindent\textbf{Attack analysis:}
We use one real case in the MCP Python SDK to illustrate the details. As we described before, MCP specifies that the ToolListChangedNotification needs to be sent from the server to the client, informing it that the list of related tools it offers has changed. Semantically, this notification acts as a change-visibility guardrail, enabling the client to remain aware of server-side changes.

From the SDK developer's perspective, this clause is optional and can not be encoded, which results in a non-implementation in the Python SDK. As shown in~\cref{fig:define-add-tool}, the SDK developer define this sending method called \texttt{send\_tool\_list\_changed}, but they do not hook them into the tool change mutator. As shown in~\cref{fig:send-tool-list-changed}, when the server adds or modifies a tool via \texttt{add\_tool}, it only updates the memory registry. Without the hooking of  However, it actually doesn't send a notification to the client, which means missing the notification guardrail. 

\begin{figure}[t]
	\centering
	\includegraphics[scale=0.50]{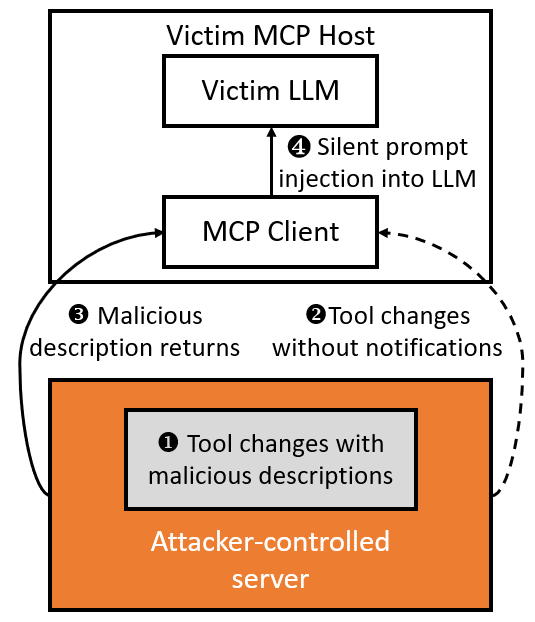}
 	\caption{Silent prompt injection.}
	\label{fig:attacking-example}
\end{figure}

An attacker can abuse this missing protection to perform prompt injection without being noticed. As shown in~\cref{fig:attacking-example}, the attacker can modify tool descriptions to embed malicious instructions. Later, when the client invokes the tool, the modified description is returned to the MCP client and then passed as context to the backend LLM. Because the notification mechanism is not implemented by the SDK, neither the victim MCP client nor the LLM is aware that the tool description has been tampered with. This results in a silent prompt-injection attack that cannot be detected from the client side. We reported this issue to the MCP security team; they acknowledged our report and stated: \emph{“We agree that servers should declare this as a best practice.”}




\noindent\textbf{Takeaway: }
Fundamentally, our finding goes beyond an ad-hoc exploiting technique, but to surface a design-level attack surface. The agent diversity feature motivates the MCP's compatibility-first design; that design legitimizes the omission of optional clauses that act as guardrails in implementations, thereby enabling compatibility-abuse attacks. Silence injection is just an instance; the root cause is the compatibility-first design of MCP. 

As a result, the resulting compatibility abuse is bidirectional: although our threat model and motivating example use a malicious server, omitting the corresponding clauses on the client side can likewise create exploitable surfaces. For expository clarity, we adopt the malicious-server case as the primary threat model in this section, while our systematic analysis in the following section covers all MCP clauses.

\section{Systematic Analysis Framework}
\label{systematic-analysis-framework}

In this section, we first discuss the existing MCP tools and their limitations. After that, we discuss the challenges and our approach for systematically analyzing the exploitable non-implemented clauses across SDKs.


\subsection{Limitations of Existing Tools}
\label{existing-tools-and-limitations}

There are several existing tools~\cite{mcp_tools_mintilify,mcp-scan,mcp-scan-web,prisma-airs,datedome-mcp,prompt-security} for discovering vulnerabilities in MCP applications. Typically, they collect a set of existing attacks as templates via heuristic methods. After that, they take the installed MCP server as input, perform a pattern-based check, and output security warnings when predefined attacking patterns are found. For example, when a malicious MCP server has prompt-injection instructions in tool descriptions, the MCP-scan tool~\cite{mcp-scan} would flag this malicious tool description and output \emph{``Prompt injection detected''}

However, these works have several shortcomings. First, these works have a limited analysis scope; they all focus on the server-level and miss the potential risks carried by non-implemented clauses across SDKs. Second, their template-based method is case-specific with expert bias and can not emulate all potential risks. For example, the mcp-scan only identifies three types of attacks: prompt injection, tool poisoning attacks, and toxic flows. It couldn't show that the root cause of the silence prompt injection is the missing of notification sending mechanism in the Python SDK. As a result, the existing tools can not solve the problem fundamentally, and we need a systematic analysis that could identify all potential risks carried by non-implemented clauses.


\subsection{Challenges and Approach Overview}
\label{challenges-and-approach}

Motivated by our root-cause analysis and the limitations of existing tools, we propose a language-agnostic, AI-enhanced approach to systematically identify those non-implemented clauses across SDKs and measure their security consequence. Our high-level idea follows the methodology behind the manual analysis discussed in~\cref{motivation}: we first systematically identify omitted clauses in concrete SDK implementations by switching perspectives from the MCP designer to the SDK developer (root-cause analysis), and then assess whether those omissions are exploitable from the attacker’s perspective (exploitation analysis).

In practice, we realize this two-phase idea as a three-step pipeline shown in~\cref{fig:mcp_architecture}. First, to make the analysis language-agnostic and reproducible across SDKs, we propose \textit{Universal IR Generator}, which normalizes heterogeneous SDKs into common intermediate representations (IR). Second, to make a scalable analyzer that doesn't rely on specific patterns with auditable analysis, we propose \textit{hybrid static-LLM analysis}, which combines the unique advantages of static analysis and LLMs. These two steps systematically and automatically implement the first phase. Third, we conduct \textit{modality-based exploitation analysis}, which abstracts universal attack modalities beyond specific templates, and identify those bugs that have potential security impacts. 

\subsubsection{Challenges}

Despite the simplicity of the three-step outline, each step poses unique challenges. For the first step, MCP has ten different official SDKs(i.e., Python, TypeScript, Go, Kotlin, Swift, Java, C\#, Ruby, Rust, PHP), and the sets keep expanding (Ruby SDK first release on May 2025~\cite{mcp-ruby-sdks}).  To our knowledge, there is no off-the-shelf IR generator that uniformly covers all these languages. Using a per-language IR generator would introduce inconsistencies and thereby hinder any unified analysis across SDKs. To overcome this challenge, we need to develop a language-insensitive IR generator. 

For the second step, even with a unified IR across SDKs, identifying all potentially non-compliant code implementations remains challenging. MCP compliance entails variability across SDKs/languages (style/idiom differences) and space breadth across clauses. Two mainstream approaches exist: pattern-based static analysis and LLM-based analysis.  Pattern-based analysis yields deterministic, reproducible findings, but faces a trade-off: too many patterns cause pattern explosion along both axes, while too few leave significant undercoverage. LLM-based methods avoid patterns via semantic reasoning, but risk hallucination and limited auditability unless their conclusions are grounded in IR evidence.

For the third step, even after we have identified non-implemented clauses across different SDKs, it remains challenging to distinguish which of them are truly exploitable from those that behave like benign code bugs. As discussed in~\cref{existing-tools-and-limitations}, validating SDKs only against a limited catalog of attack templates is heuristic and heavily depends on experts’ domain knowledge. Moreover, template-based checking is restricted to known attack patterns and cannot systematically uncover novel, future attack vectors.


\subsubsection{Approach Overview}
\label{approach-overview}

\begin{figure*}[ht]
	\centering
	\includegraphics[scale=0.55]{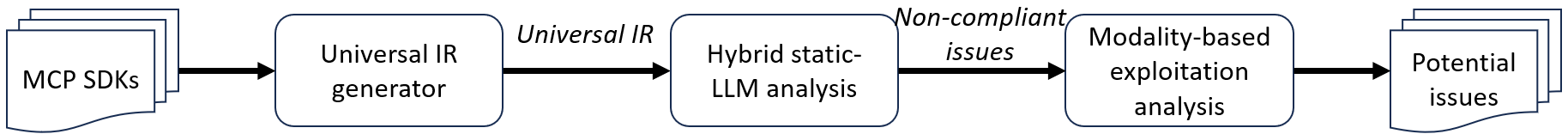}
	\caption{The architecture of our approach.}
	\label{fig:analyze-tool-overview}
\end{figure*}




For universal IR generation, our key insight is that the semantic essence of an MCP clause is performing a specific action under certain conditions. In code implementation, this corresponds to when the condition is satisfied, there exists a reachable path that invokes the action functions (calls). This essence is not language or SDK-specific. Operationally, rather than using multiple language-specific IR generators with divergent IR shapes. We normalize the code of each SDK into a unified conditional call list and a function-definition catalog. This per-SDK normalization provides us a basis for further unified analysis across different-language SDKs (see~\cref{IR-generation}).


For cross-SDK/clauses compliance analysis, our key insight is that the static analysis slices the program via call/definition graphs, reducing the search space from the full code space to a finite subgraph. Within this constrained space, an LLM performs semantic reasoning about condition satisfaction and the functional intent of the invoked function. As a result, a static analysis framework can significantly mitigate the hallucinations from whole-code LLM reasoning. For each slice, an LLM-based checker significantly avoids pattern explosion across MCP clauses and languages (see~\cref{compliance-analysis}).

For exploitable analysis, our key insight is that a typical clause specifies when certain actions are permitted under what conditions. It can be abstracted as two-dimensional: payload guardrail (action) and timing guardrail (conditions). If an attacker wants to exploit the non-implementation, the missing guardrail must allow the attacker to control at least one dimension. If none of them can be controlled, it is merely a normal bug. This abstract is an attacking technique unrelated; it eliminates the burden of maintaining an attack set and new technique adoptable. By taking the Cartesian product across these two dimensions, we build three attacking modalities (payload-only, timing-only, and joint). And build a modality-based pipeline to uncover those potential risks(see more details in~\cref{weaponize-analysis}).



\subsection{Universal IR Generation}
\label{IR-generation}

\begin{algorithm}[t]
\DontPrintSemicolon
\SetKwFunction{LoadParser}{LoadParser}
\SetKwFunction{Parse}{Parse}
\SetKwFunction{Matches}{Matches}
\SetKwFunction{FindCondition}{FindCondition}
\SetKwFunction{WriteJSONL}{WriteJSONL}
\KwIn{SDK repositories $\mathcal{R}$; for each language $lang$: parser $P_{lang}$ and query packs $Q^{call}_{lang}$, $Q^{def}_{lang}$}
\KwOut{For each $lang$: \texttt{calls\_<lang>.jsonl}, \texttt{defs\_<lang>.jsonl}}

\ForEach{$R \in \mathcal{R}$}{
  $lang \leftarrow \texttt{DetectLanguage}(R)$\;
  $P \leftarrow$ \LoadParser{$lang$}\;
  \texttt{CallList[}$lang$\texttt{]} $\leftarrow [\,]$;\quad \texttt{DefCatalog[}$lang$\texttt{]} $\leftarrow [\,]$\;
  \ForEach{file $p$ in $R$}{
    $T \leftarrow$ \Parse{$P, p$}\;

    $C \leftarrow \Matches(Q^{call}_{lang}, T)$;\quad
    $D \leftarrow \Matches(Q^{def}_{lang},  T)$\;

    \ForEach{$m \in C$}{
    $cond \leftarrow$ \FindCondition{$m$}\;
      append $\{kind, name, recv, cond, \texttt{filepath}(m)\}$ to \texttt{CallList[}$lang$\texttt{]}\
    }
    
    \ForEach{$m \in D$}{
      append $\{kind, name, params, \texttt{filepath}(m)\}$ to \texttt{DefCatalog[}$lang$\texttt{]}\;
    }
  }
  \WriteJSONL{\texttt{calls\_\$lang\$.jsonl}, \texttt{CallList[}$lang$\texttt{]}}\;
  \WriteJSONL{\texttt{defs\_\$lang\$.jsonl},  \texttt{DefCatalog[}$lang$\texttt{]}}\;
}

\caption{Language-Agnostic IR Construction}
\label{alg:ir-construction}
\end{algorithm}

Following the first insight in~\ref{challenges-and-approach}, our IR only needs to capture the function calls and related specific conditions. This information is already explicit at the AST level: function calls appear as call nodes, and their guarding conditions are represented by surrounding guard conditions( e.g., if/loop/cases). Therefore, we could first parse each SDK with its language-specific parser to obtain an AST, and extract calls with guard and function definitions from the AST and assemble them into a unified, language-agnostic IR.




More specifically, we design an algorithm to implement this. As shown in~\cref{alg:ir-construction}, for each SDK, we first load the language-specific parser (line 3) and parse every file to obtain a language-specific AST (line $6$). After that, we capture language-internal manifestations of function calls\&conditions and definitions via predefined per-language query packs. Third, we perform normalization for call set C and definitions set D (line 7), reshaping language-specific into a unified record schema (line 10, 13). Finally, we materialize the IR by emitting calls\_<lang>.jsonl and defs\_<lang>.jsonl (line 16,17). This makes normalization intrinsic to extraction, abstracting away both code-level and AST-level language idiosyncrasies, and yielding a language-agnostic, condition-aware IR ready for compliance analysis.

For example, considering the \texttt{ToolListChanged} notification mechanism in Python and TypeScript SDK. In Python SDK, the parser generates the definitions of \texttt{send\_tool\_list\_changed} method. However, it doesn't found the conditional caller because the \texttt{add\_tool} doesn't call it. For TypeScript SDK, the source code is shown in~\cref{fig:typescript-hook}. The parser generates the definitions of \textbf{sendToolListChanged} methods, and captures its conditional caller \texttt{\_creatingRegisteredTool} in \texttt{update: (updates)} cases. It appears in the IR as a case-condition call.


\begin{figure}[t]
\small
\centering
\begin{subfigure}{0.45\textwidth}
\begin{minted}[fontsize=\scriptsize,highlightlines={1,3},breaklines=true, samepage]{c}
sendToolListChanged() {
    if (this.isConnected()) {
        this.server.sendToolListChanged();
    }
}
\end{minted}
\caption{The definition of sendToolListChanged method.}
\label{fig:send-tool-list-changed-TypeScript}
\end{subfigure}

\vspace{10pt}
\begin{subfigure}{0.45\textwidth}
\begin{minted}[fontsize=\scriptsize, breaklines=true, highlightlines={5,9,10}, samepage]{c}
private _createRegisteredTool(
  ...
): RegisteredTool {
    const registeredTool: RegisteredTool = {
        ...
        update: (updates) => {
            ...
            if (typeof updates.enabled !== "undefined") registeredTool.enabled = updates.enabled
            this.sendToolListChanged()
        },
    };
    ...
}
\end{minted}
\caption{The definition of \_createRegisteredTool method.}
\label{fig:define-add-tool-Type-Script}
\end{subfigure}
\caption{The tool modification hook in TypeScript SDK.}
\label{fig:typescript-hook}
\end{figure}


\subsection{Hybrid Static-LLM Analysis}
\label{compliance-analysis}



Following the second insight in~\cref{challenges-and-approach}, we could anchor LLM reasoning on IR slices rather than the entire code base. This IR $\rightarrow$ source-slice keeps the search space narrow and auditable, and still allows cross-SDKs/clauses generalization without pattern explosion.



More specifically, we employ an IR-first, self-refining loop. As shown in~\cref{alg:ir-refine}, the LLM first generates keywords from the rule semantics (line 2) and uses them to query the condition-aware IR (line 4). After finding the related conditional calls and function definitions, it reads only the corresponding source-code slices (callee definitions and call sites) to perform semantic reasoning and produce a confidence score (line 7). If this confidence is insufficient, we let the LLM refine its understanding of the clause semantics and generate new keywords (line 12) to search the IR again; the LLM then performs another round of reasoning and produces an updated confidence score. The LLM stops reasoning when the confidence meets or exceeds a threshold $\tau$ (line 8). If it still fails to reach $\tau$ after the maximum number of iterations $M$, we fall back to a broad search over the entire codebase as a last resort (line 14). This “IR $\rightarrow$ source-slices $\rightarrow$ self-refining $\rightarrow$ full-source fallback” procedure makes every retrieval step IR-validated, keeps the search space narrow and auditable, and still supports cross-SDK and cross-language semantic generalization.

\begin{algorithm}[t]
\DontPrintSemicolon
\caption{IR-Driven Self-Refining Loop}
\label{alg:ir-refine}
\SetKwFunction{QueryIR}{QueryIR}
\SetKwFunction{ReadSlices}{ReadSlices}
\SetKwFunction{LLMReason}{LLMReason}
\SetKwFunction{Refine}{Refine}
\SetKwFunction{Paths}{Paths}
\KwIn{Rule set $\mathcal{U}$; IR files (\texttt{defs.jsonl}, \texttt{calls.jsonl}); source repo $S$; threshold $\tau$; max iters $M$}
\KwOut{\texttt{Result}[\,$r$\,] for each $r \in \mathcal{U}$: \{\texttt{status}, \texttt{evidence}, \texttt{files\_analyzed}, \texttt{confidence}\}}

\ForEach{$r \in \mathcal{U}$}{
  $E \leftarrow [\,]$;\quad $q \leftarrow 0$;\quad $K \leftarrow$ \Refine($r$, \textit{init})\;

  \For{$i \leftarrow 1$ \KwTo $M$}{
    $H \leftarrow$ \QueryIR(\texttt{defs.jsonl}, \texttt{calls.jsonl}, $K$)\;
    $S_i \leftarrow$ \ReadSlices($S$, $H$)\;
    $E \leftarrow E \cup S_i$\;
    $(J, q) \leftarrow$ \LLMReason($r$, $E$)\;

    \If{$q \ge \tau$}{
      \texttt{Result}[$r$] $\leftarrow$ \{\texttt{status}: $J$, \texttt{evidence}: $E$, \texttt{files\_analyzed}: \Paths($E$), \texttt{confidence}: $q$\}\;
      \textbf{continue}\;
    }

    $K \leftarrow$ \Refine($r$, $E$, $J$)\;
  }

  $(J, q) \leftarrow$ \LLMReason($r$, $S$)\;

  \texttt{Result}[$r$] $\leftarrow$ \{\texttt{status}: $J$, \texttt{evidence}: $E$, \texttt{files\_analyzed}: \Paths($E$), \texttt{confidence}: $q$\}\;
}
\end{algorithm}


For example, consider the clause \emph{``When the list of available tools changes, servers that declared the `listChanged` capability SHOULD send a notification''}. For the TypeScript and Python SDK, the LLM reveals two different notifiers (i.e., \texttt{sendToolListChanged}, \texttt{send\_tool\_list\_changed}) that match the rule, even though their names are different. Following IR edges, the LLM opens the callee/call-site source code slides. For the TypeScript SDK, the IR reveals that a notifier \texttt{sendToolListChanged} is called in the source file when the conditions are satisfied. The LLM marks the TypeScript SDK as compliant. For Python SDK, although the reveals the notifier \texttt{send\_tool\_list\_changed}. It finds no call sites that reach it, and marks the Python SDK non-compliant.



\subsection{Modality-based Exploitability Analysis}
\label{weaponize-analysis}

\begin{figure}[t]
\centering
\begin{tikzpicture}[>=Stealth, x=0.9cm, y=0.9cm]
  \draw[->, thick] (-3,0) -- (3.4,0)
    node[below right]{\textbf{\small Time control}};
  \draw[->, thick] (0,-2.4) -- (0,2.7)
    node[above left]{\textbf{\small Payload control}};
  \node at ( 1.9, 1.6) {\small PyTy,  both-controllable}; 
  \node at (-1.9, 1.6) {\small PyTn,  payload-only}; 
  \node at ( 1.9,-1.6) {\small PnTy, timing-only}; 
  \node at (-1.9,-1.6) {\small PnTn, non-exploitable}; 
\end{tikzpicture}
\caption{Attack modality space.}
\label{fig:attack-modality-quadrant}
\end{figure}
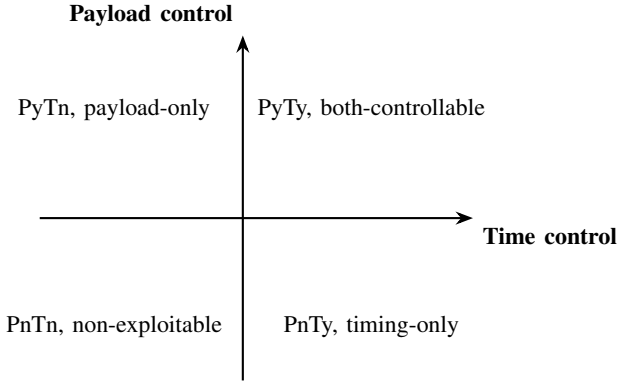

Following the third insight in~\cref{challenges-and-approach}, the absence of a clause permits three attack modalities. As shown in~\cref{fig:attack-modality-quadrant}, the first is \emph{payload-and-timing controllable}: the clause acts as a guardrail on both when and what to send, so missing such a clause lets the attacker send malicious content to the LLM whenever they want (PyTy in~\cref{fig:attack-modality-quadrant}). Second, the clause acts primarily as a timing restriction, while the payload is still protected by other implemented items. As a result, the attacker cannot control the message content, but can control when it is sent, which admits multiple attack semantics such as DoS. Third, the attacker cannot control the sending time, but can control the malicious content being sent (payload-only). In contrast, if the attacker can control neither payload nor timing, the missing clause has no attack semantics in our model and is treated as just a normal bug.


We implement another LLM-based pipeline to proceed with this exploitable analysis. For each MCP clause, we first execute a semantic analysis to identify the sender, message content, and sending status. Based on this, we infer that this clause is a timing or payload restraint. Consequently, if this constraint is removed, the sender can abuse the sending time or message payload and infer potential attack vectors.

We still use the MCP clause \emph{``When the list of available tools changes, servers that declared the `listChanged` capability SHOULD send a notification''}. Semantic analysis yields that the sender is the server, the message is ToolListChangedNotification, and the timing is a tool change. This clause is a timing restraint because it restrains when the notification is sent. If this wiring is omitted, a malicious server can both \emph{modify} tool description (payload) and \emph{choose when} such changes become invisible (timing) by muting the notification. Making a normal prompt injection become silence prompt injection.


\section{Evaluations}
%


Our approach has three steps: universal IR generation, hybrid static-LLM analysis, and exploitability analysis. The result of IR generation is just a pre-processing artifact for serving the second step. In contrast, the result of hybrid static-LLM analysis demonstrates the potential non-implementations across SDKs, and the result of exploitability analysis keeps those real potential risks. As a result, we mainly conduct experiments from two aspects: non-implementations evaluation and exploitability analysis evaluation.


\subsection{Non-implementations Evaluation}


\subsubsection{Evaluation Setup}
We run our clause non-implementation analyzer on all \textbf{ten official MCP SDKs} (\textit{ten programming languages, 275 protocol clauses}) to demonstrate its accuracy, scalability, and cost-efficiency. 
To our knowledge, this is the first systematic clause-level implementation analysis spanning the entire MCP specification across multi-language SDKs, i.e., at full protocol scale. 
We treat the analyzer like a static program analysis tool: after translating each SDK into our IR and applying the LLM-based checker. As a result, we use a set of classic static analysis criteria (i.e., false positive, false negative, precision, recall) to measure the accuracy of our tool. 
This principled approach (rather than ad-hoc success metrics) ensures the evaluation is reproducible and unbiased, aligning with established static-analysis methodology. 
We also assessed the analyzer’s cost and runtime, logging the total monetary cost and wall-clock time per SDK and per rule, to gauge practical usability. 
The focus of our evaluation is to confirm that the tool achieves high recall (minimal missed non-implemented issues) and high precision (low false alarms) cost-effectively.

\subsubsection{Raw Findings}

\autoref{tab:raw-findings} summarizes the raw findings produced by our analysis for each SDK. 
Out of 275 MCP clauses, each SDK was found to omit a substantial subset. 
On average, roughly \textbf{30–60\% of protocol clauses per SDK} were flagged as non-compliant, confirming that a noticeable fraction of intended behaviors are missing in practice. Based on each non-implemented clause, we launch the exploitability analysis. The total number of exploitable clauses is very close to non-implementations per SDK, showing that most of the non-implemented clauses are potential risks rather than normal bugs.

Crucially, the vast majority of these non-implemented instances involve optional or conditional clauses (“SHOULD”, “MAY”, or conditional “MUST” in the specification) rather than unconditional requirements. 
For example, the PHP SDK missed 202 clauses (169 of them optional), whereas the Python SDK missed 93 (76 optional), in contrast to only 17–42 violations of mandatory “MUST” clauses in those SDKs. 

Such results empirically reinforce our key observation (\S \ref{motivation}) that MCP’s high compatibility derived from its many optional clauses is the principal driver of these “non-compliant” cases, rather than willful disregard of required clauses. 
In other words, SDK developers tend to be in compliance with the core mandatory protocol rules but often ignore implementations of those optional features, creating a broad attack surface consistent with MCP’s compatibility-by-design philosophy. 
These raw findings establish that our tool scales across all 10 languages and successfully surfaces the intended clause-level gaps for each, providing a first-ever cross-SDK view of MCP implementation coverage.

\begin{table}[t]
\centering
\caption{Raw implementation findings across SDKs. \textit{I} refers to implemented. \textit{NI} refers to non-implementations. \textit{NI-O} refers to non-implementation of optional clauses. \textit{NI-NO} refers to omission of non-optional implementation. }
\label{tab:raw-findings}
\scriptsize
\begin{tabular}{lccccc}
\toprule
\textbf{SDKs} & \textbf{I} & \textit{\textbf{NI-O}} & \textit{\textbf{NI-NO}} & \textbf{NI} & \textbf{Exploitable} \\
\toprule
Python     & 182 & \textit{76} & \textit{17} & 93  & 93 \\ 
TypeScript & 193 & \textit{62} & \textit{20} & 82  & 81 \\ 
Go         & 195 & \textit{62} & \textit{18} & 80  & 80 \\ 
Kotlin     & 128 & \textit{111} & \textit{36} & 147 & 147 \\ 
Swift      & 104 & \textit{135} & \textit{35} & 170 & 170 \\ 
Java       & 169 & \textit{82} & \textit{24} & 106 & 106 \\ 
C\#        & 177 & \textit{76} & \textit{22} & 98  & 96 \\ 
Ruby       & 108 & \textit{132} & \textit{35} & 167 & 167 \\ 
Rust       & 159 & \textit{90} & \textit{26} & 116 & 115 \\ 
PHP        & 73  & \textit{169} & \textit{42} & 211 & 210 \\ 
\toprule
\textbf{Total} & \textbf{1588} & \textit{1015} & \textit{275} & \textbf{1270} & \textbf{1265} \\
\bottomrule
\end{tabular}
\end{table}

\subsubsection{Precision and Recall}



We evaluated the accuracy of the tool’s findings in terms of false positives (cases where a detected non-implemented is actually implemented correctly) and false negatives (missed non-implementations). 
We followed a rigorous validation process: for each SDK, we randomly sampled 20 flagged (non-compliant) clauses and 20 unflagged (compliant) clauses, then performed a human review with three experts, including two reviewers and one adjudicator, to determine the ground truth. 
In total, this amounted to 400 judgments (20×2×10) and over 800 decision points, including adjudication, ensuring statistically meaningful and unbiased accuracy metrics~\cite{conroy2015sample}. As shown in \autoref{tab:fp-fn}, the tool achieves high precision and recall overall. 
False-positive rates remained under 35\% in all SDKs, and false-negative rates under 30\%. 
On average, the false-positive rate was ~14\% and the false-negative rate ~13.5\%, corresponding to approximately 86\% precision and 87\% recall across the 10 SDKs. 
In fact, in 8 of 10 SDKs, the analyzer attained at least 85\% recall, and in half of the SDKs, it achieved 90–100\% precision (zero or near-zero false alarms). 
The few outlier cases (e.g., the Swift SDK saw ~30\% FN rate, and Python ~35\% FP rate) reflect challenging corner cases, but even there the tool correctly identified the majority of issues. 
No SDK had an FN rate above 30\%, meaning the analyzer caught at least 70\% of all clause violations in the worst case, and much higher fractions in most cases. 
This level of recall is critical for a security-oriented analyzer, as it means very few true non-implementation issues go undetected. 
Meanwhile, a precision around 85–90\% indicates that the findings are largely actionable, with minimal noise. 
These results demonstrate that our LLM-guided static analysis approach can reliably pinpoint clause-level implementation flaws across disparate codebases with accuracy, despite the complexity of multiple languages.

\begin{table}[t]
\centering
\caption{Detection performance across SDKs (20 cases each)}
\label{tab:fp-fn}
\small
\begin{tabular}{lccccccc}
\toprule
\textbf{SDK} & \textbf{TP} & \textbf{FP} & \textbf{TN} & \textbf{FN} &
\textbf{Precision} & \textbf{Recall} \\
\toprule
Python     & 13 & 7 & 19 & 1 & 65.0\% & 92.9\% \\
TypeScript & 15 & 5 & 17 & 3 & 75.0\% & 83.3\% \\
Go         & 18 & 2 & 18 & 2 & 90.0\% & 90.0\% \\
Kotlin     & 18 & 2 & 17 & 3 & 90.0\% & 85.7\% \\
Swift      & 20 & 0 & 14 & 6 & 100.0\% & 76.9\% \\
Java       & 18 & 2 & 18 & 2 & 90.0\% & 90.0\% \\
C\#        & 17 & 3 & 18 & 2 & 85.0\% & 89.5\% \\
Ruby       & 18 & 2 & 18 & 2 & 90.0\% & 90.0\% \\
Rust       & 18 & 2 & 19 & 1 & 90.0\% & 94.7\% \\
PHP        & 17 & 3 & 15 & 5 & 85.0\% & 77.3\% \\
\bottomrule
\end{tabular}
\end{table}

\subsubsection{Cost Effectiveness} 
We further examine the cost and scalability of our approach. Using a commercial LLM API, the total cost to analyze all 275 clauses across 10 SDKs (2,750 clause checks) was only \$541.87. 
This works out to \$54.2 per SDK on average, or about \$0.20 per clause check. 
In practical terms, each individual clause's non-implemented decision costs only a few dimes, which is an extremely low cost for a thorough static audit of an SDK’s adherence to 275 rules. 
Given the accuracy discussed above, this translates to an estimated cost well under \$1 per true issue detected (e.g., ~\$0.6 per confirmed clause violation, assuming ~86\% of the 1002 flagged issues are true positives). 
We consider this cost–benefit ratio highly favorable for deployment: for roughly the price of a single engineer-hour, our tool can audit an entire SDK and find dozens of potential vulnerabilities. 
The approach is also time-efficient, which means that running the analyzer on one SDK (275 rules) completes in a matter of minutes, easily fitting into CI pipelines or regular security testing. 
In summary, the monetary and computational overheads are low enough to make our implementation analysis usable in practice, even at scale. 
We emphasize that this cost-effective scalability is a direct consequence of our design: by leveraging a universal IR and automating analysis with an LLM, we can validate thousands of clause-SDK pairs quickly and cheaply, given that this would be prohibitively labor-intensive to do manually. 
Overall, the non-implementation analysis evaluation shows that our tool is effective (high recall/precision), scalable (10 SDKs, 275 clauses), and practical (affordable automated analysis).

\subsection{Exploitable Analysis Evaluation}

\subsubsection{Ethical Consideration}
For ethical considerations, we intend to evaluate the result of exploitable analysis without causing real-world impact. Further, all potential risks should be responsibly reported to the relevant communities through proper channels. And we should only discuss acknowledged ones.

For the environment setup, we conduct a separate agent testbed fully controlled by ourselves. We install the official MCP SDKs locally, create self-controlled malicious agents to launch all attacks. The victim client and LLM are both created by our own account. No commercial or public agents are exploited.

For responsible disclosure, we reported our findings through two channels: First, we reported through the MCP official private HackerOne security channel~\cite{mcp_security}. Second, we reported SDK-related issues to the corresponding communities. As the second channel is public, we want to avoid any potential risks being exploited. We intentionally redacted any exploitation methods and reported this problem as a non-implementation issue. In this paper, we discuss only a small set of acknowledged cases that cover our three-attack modalities. We don't show any exploitation details. All unconfirmed issues will not be discussed until community acknowledgment.

\subsubsection{Scalable Responsible Disclosure}
\label{sec:scalable-disclosure}

Our analyzer surfaced 1,265 potential risks, making one-by-one triggering, reporting and confirmation infeasible at scale. For our researchers, preparing and submitting an individual report takes approximately 30 minutes, implying more than 500 hours in total. For community maintainers, triaging and responding to each case typically takes about two days on average. When a large number of reports arrive in a short window, limited reviewers and time make it impractical to review and respond to every report individually within that cadence. Consequently, we need a new disclosure method that is responsible while keeping timelines and workload controllable.

To achieve this, we adopt a case-driven, CI/CD method. First, we submit a small, manually-written subset of potential issues to a specific SDK developer who is not only a specific SDK developer but MCP community maintainer. Second, after they validate that these cases are real, we told them that these issues are not ad hoc but identified by our approach to spark the community's interest. Building on that engagement, our tool is invited as part of the MCP community's performance testing workflow, and the tool-identified potential risks become the test cases for performance testing and fixing the pipeline. This shifts disclosure from one-by-one, SDK-specific manual reporting to a continuous, maintainer-executable verification and remediation process—across SDKs, helping raise the overall security posture of the MCP community.

\subsubsection{Seeded manual reports}

We present three representative, acknowledged cases to illustrate how specification non-implementations can translate into potential security risks. These cases were hand-reported across different SDKs and are selected to cover the three attack modalities introduced in Section~\ref{weaponize-analysis}. Exploitation details are hidden for ethics.

\begin{figure}[t]
\small
\centering
\begin{subfigure}{0.45\textwidth}
\begin{minted}[fontsize=\scriptsize,highlightlines={4,5,6},breaklines=true, samepage]{c}
def is_token_valid(self) -> bool:
    """Check if current token is valid."""
    return bool(
        self.current_tokens
        and self.current_tokens.access_token
        and (not self.token_expiry_time or time.time() <= self.token_expiry_time)
    )
\end{minted}
\caption{The definition of is\_token\_valid method.}
\label{fig:is-token-valid}
\end{subfigure}

\vspace{10pt}
\begin{subfigure}{0.45\textwidth}
\begin{minted}[fontsize=\scriptsize, breaklines=true, highlightlines={3,6,9}, samepage]{c}
async def async_auth_flow(...)
    ...
    if self.context.is_token_valid():
        self._add_auth_header(request)

def _add_auth_header(self, request: httpx.Request) -> None:
    """Add authorization header to request if we have valid tokens."""
    if self.context.current_tokens and self.context.current_tokens.access_token:
        request.headers["Authorization"] = f"Bearer {self.context.current_tokens.access_token}"
\end{minted}
\caption{The method of setting authorization header.}
\label{fig:set-authorization-header}
\end{subfigure}
\caption{The token checking method in Python SDK.}
\label{fig:token-check-omission}
\end{figure}

First, we present one Python SDK issue for payload-only modality. The MCP specification requires that \emph{``MCP clients MUST NOT send tokens to the MCP server other than ones issued by the MCP server's authorization server.''} In the current Python SDK, the client checks only token presence and expiry before attaching the Authorization header. And it does not verify that the token was issued by the authorization server of the connected MCP server. As illustrated in \cref{fig:token-check-omission}, \texttt{is\_token\_valid} returns true when a token object exists and is unexpired, but it never validates that tokens were issued by the expected authorization server before they are used. As a result, the token intended for one server may be sent on another. This grants the attacker control over the payload: when receiving the token, they can send malicious requests to unintended servers and execute the attacker-controlled requests.

We submit this report to the Python SDK developer as a non-compliant report. They acknowledge our findings by adding  \emph{ready for work} tag. Besides, they also added \emph{P0} (priority 0), \emph{PR welcome}, \emph{good first issue} tag. These signals indicate that our reports are taken seriously by the community even without exploitation details, underscoring the effectiveness of our tool’s analysis.

Second, we present another Python SDK issue for the time-only controllable modality. The MCP specification requires that \emph{``The frequency of pings SHOULD be configurable.''} However, the MCP Python SDK provides no mechanism for users to configure and limit ping behavior. As shown in~\cref{fig:ping-rate-limitation-omission}, the client/server can send/receive pings without configurable interval limits, creating a potential DoS vector. Although the ping format has been fixed, the attacker from both client/server can send repeated pings to exhaust responders' resources. We reported this as a non-implementation finding; maintainers acknowledged it by labeling the issue \emph{ready for work}, \emph{P3}, and \emph{improves spec compliance}.

\begin{figure}[t]
\small
\centering
\begin{subfigure}{0.45\textwidth}
\begin{minted}[fontsize=\scriptsize,highlightlines={4,5,6},breaklines=true, samepage]{c}
 async def send_ping(self) -> types.EmptyResult:
    """Send a ping request."""
    return await self.send_request(
        types.ServerRequest(types.PingRequest()),
        types.EmptyResult,
    )
\end{minted}
\caption{The definition of send\_ping method of server side.}
\label{fig:send-ping}
\end{subfigure}

\vspace{10pt}
\begin{subfigure}{0.45\textwidth}
\begin{minted}[fontsize=\scriptsize, breaklines=true, highlightlines={3,5,6}, samepage]{c}
 async def _received_request(...) -> None:
    ...
    case types.PingRequest():
        with responder:
            return await responder.respond(types.ClientResult(
            root=types.EmptyResult()))
\end{minted}
\caption{The ping request handling method of the client.}
\label{fig:ping-response}
\end{subfigure}
\caption{The ping rate limitation in Python SDK.}
\label{fig:ping-rate-limitation-omission}
\end{figure}




Third, we present another Python SDK issue for payload\&timing controllable modality. Similar to the silence injection we discussed in~\cref{motivation}, there are also other optional notifications (e.g., ResourceListChangedNotification, PromptListChangedNotification) specified but not implemented by the Python SDK. Similarly, the attacker can exploit these silent injections to send malicious instructions (payload) to the LLM with notification muted. These gaps share the same root cause as \texttt{ToolListChangedNotification}, and we report them following the same HackerOne report channel as the tool-notification report. They also confirm these issues.

\subsubsection{Community Impact}
\label{community-impact}

We present community-side evidence that validates the disclosure approach described in Section~\ref{sec:scalable-disclosure}: maintainers confirmed our findings, expressed sustained interest, and invited tool integration.

For manual reports/confirmations, to date, we have reported 3 issues to MCP Hackerone channel, 20 issues to the Python SDK developer, and 2 to Go and 1 to TypeScript SDK developer. For the Hackerone channel, all three reports are confirmed. For 20 Python SDK report, 15 of them are acknowledged by assigning a \emph{ready for work} tag from the maintainers. Especially, 5 were triaged as high priority (3 P0 and 2 P1). Besides, Go and TypeScript developers also confirm one of our reported issues separately. Given the number of potential risks identified, reporting and acknowledgment are ongoing, and these counts are expected to increase if we continue.

Our reports raise MCP community's interest, after these manual reports submission, the MCP maintainer say : \emph{``A lot of reported issues go very much in line with a SEP we currently have open for Conformance testing''}. And, they also say: \emph{``It would be great if you'd be able to contribute to that implementation with runnable tests that demonstrate the compliance failure''}. These all shows that they are interested in adopting our approach.

After communicating with them and showing our intention to open source our tool, they say: \emph{``We'll want to formalize the processes for conformance testing which we'll want to have a vote on since it will affect future spec changes and all SDKs. Rough plan is to get working on this in a separate repo, and update this issue with design decisions as we go.''} All these communications show our tool is not only a toy example but with promising community impact.

\section{Discussions}
\label{discussion}

\subsection{Mitigation Discussion}
\label{mitigation-discussion}
According to our communications with MCP developers and communities, we propose three mitigations for different aspects: agent developers, MCP SDK developers, and MCP designers.

\noindent\textbf{For agent developers}: As discussed in~\cref{motivation}, many agents are built on top of official SDKs whose optional clauses are not implemented. A malicious server/client could about the missing guardrails to exploiting multiple techniques and attack the LLM. As a result, the agent developer should not assume these guardrails exist in SDKs. Treat third-party MCP servers/clients as potentially malicious, and use only server/clients that have explicitly audited against our omission list, and reject those that use mechanisms that rely on non-implemented clauses. For example, if a server using Python SDK needs to change its tools dynamically, it should be treated as potentially malicious because of the potential risks of silent prompt injection. In short, agent developers should only use audited and trusted third-party MCP servers/clients.

\noindent\textbf{For MCP SDK developers: }
As discussed in~\cref{community-impact}, the MCP community is considering integrating our analyzer into the standard engineering process. As a result, the SDK maintainer can use our approach-generated omission list as a fixing checklist for current releases. Whenever new clauses or features are added, they could also run our analyzer as a pre-merge gate, and block merges if required guardrails (e.g., \texttt{Tool/Prompt/ResourceListChanged}, token issuer checks, ping configurability) are not detected in IR. In practice, these two adoption scenarios could ensure that the clause is properly implemented in the current version, and the implementation is present before version updating.

\noindent\textbf{For MCP designers.}
We appreciate the MCP community’s willingness to integrate our analyzer into the SEP pipeline; this will substantially improve clause-to-implementation traceability and raise the security bar across SDKs.
Beyond tooling, we recommend the designer to adopt our \emph{designer $\rightarrow$ developer $\rightarrow$ attacker} perspective shift: reason from clause omissions to their security impact, and promote a minimal \emph{security baseline} of high-impact clauses from SHOULD/OPTIONAL to MUST.

Besides, when introducing new clauses, they can also run this perspective-shift security audit: they could evaluate the implementations in official SDKs, and evaluate exploitability from the attacker's view before standardizing the new clause into the existing specification. In short, a small, explicit security baseline and a perspective-shifting review can shrink the compatibility-abuse surface while preserving MCP's high compatibility.

\subsection{Limitation Discussion}
\label{limitation-discussion}

\noindent\textbf{Reliability of our tool: }
We acknowledge that due to language idioms and continuously evolving new SDKs and new clauses, our analyzer could carry over-approximate analysis results and may miss complex, language-specific clause implementations of certain clauses. This is a deliberate trade-off for the first systematic analyzer across multiple SDKs: we favor a universal, IR-guided, cross-SDK/clauses identification of potential omissions, over building a per-language, highly specialized analyzer. Although the latter could capture more idioms in a given language and clause, it would sacrifice cross-SDK/clause generality and would not scale as new SDKs/clauses arrive in a fast-moving ecosystem. 

For example, within roughly one year, the MCP specification shipped five substantive versions (2024-10-07, 2024-11-05, 2024-11-05-final, 2025-03-26, 2025-06-18) and continued with a draft thereafter. In the meantime, new official SDKs are onboarded in the same window (e.g., Ruby in May 2025). A per-language analyzer would require continuous re-engineering for each revision and each SDK, undermining scalability. As future work, a hybrid design is feasible: keep the IR-first core for scalability, and optionally add language-specific plugins to reduce false negatives without losing cross-SDK compatibility.

\noindent\textbf{Why we do not report every case: }We don't report all cases because of another trade-off: The number of potential issues identified by our approach is large, whereas maintainer bandwidth is limited. According to the communications with maintainers, we prioritized integrating our tool into the performance-test SEP rather than filing numerous standalone reports.

For example, among the $20$ Python SDK issues we initially reported, maintainers manually acknowledge and label $15$ of them. After that, they invite our tool into SEP, and mark the remaining 5 items as \emph{``duplicated with SEP''}. This reflects a conscious triage decision rather than a judgment that our findings are unreal. After integration, our approach's identified results naturally become inputs to performance testing, providing continuous coverage with lower maintainers' overhead compared to a long tail of individual bug reports that maintainers may not have time to read. As a result, tool integration yields a community-driven and more sustainable impact than attempting to disclose every instance individually.


\section{Related Work}

\subsection{Attacks in LLM Agents}
Large language model agents have surged in popularity, enabling diversity tools development via a set of frameworks and tools~\cite{yao2022react,mavroudis2024langchain, openai2023chatgptplugins, yang2023auto, nakajima2023babyagi}. In the meantime, multiple papers and real-world exploits~\cite{wang2024poisoned,iqbal2024llm, unit42_paloaltonetworks,cve2023-46229} show that current LLM agents often lack robust built-in guardrails. Specifically, in the MCP scenario, several exploitation works~\cite{pillar2024mcpsecurity} show that existing techniques (e.g., prompt injection, token theft, etc.) can be exploited to make multiple attacks.

From the application aspect, the rapid integrations of LLMs into applications have prompted extensive study of their security and privacy implementations. Multiple surveys~\cite{yao2024survey,wang2025large,liu2024automatic,wang2025large} catalog the emerging threats. Specifically, LLM apps operate through natural language prompts that blur the line between code and data. This creates multiple novel injection channels analyzed by multiple works~\cite{liu2023prompt, owasp2025llmtop10, yeo2025multimodal, hung2024attention}.


While prior studies have examined AI agents and MCP mostly from the attacker’s perspective, they are largely case-specific and focus on the agent itself, while leaving the uneven implementation of clauses across SDKs as an ignored attack surface. Our contribution is motivated by a perspective shift from \emph{designer} $\rightarrow$ \emph{developer} $\rightarrow$ \emph{attacker}: we not only point out the diversity feature of agents that motivates the MCP's high-compatibility design, and systematically identify those non-implemented clauses which can be exploited to launch multiple attacks. In other words, we move beyond case-specific attack analysis and reveal a new attack surface intrinsic to MCP’s design.

\subsection{Defense Methods for Agents}
As the community grapples with securing agents, a promising avenue is the use of formal methods and principled safety frameworks to enhance ad-hoc defenses. Researchers propose a set of guard agents~\cite{xiang2024guardagent, chen2025shieldagent} with specified rules to monitor the primary agent and intervene to block out-of-policy actions. Besides, another branch of research integrates formal verification into the agent’s own decision-making loop. works such as VeriGuard, AgentHarm~\cite{miculicich2025veriguard,crouse2023formally, andriushchenko2024agentharm} propose a set formal verification method to ensure the agent adhered to the desired patterns and could even improve task success while avoiding policy violations. 

However, these defenses concentrate on the agent’s internal policy loop and overlook risks stemming from the agent’s diverse external services via MCP. In contrast, by systematically analyzing SDK-enforced clauses that constrain interactions with external services, we identify a rich set of clause-implementation omissions and show how they can be exploited via three modalities. Crucially, these omissions operate beyond the agent’s reasoning loop and are orthogonal to policy-level defenses. In other words, such approaches can not fundamentally solve the problem.

\subsection{Limitations of Existing MCP Tools}
Corresponding with the widespread adoption of MCP, there is also a set of tools to identify and mitigate the risks. MCPScan and its web is a security scanning tool that inspects MCP servers for common vulnerabilities~\cite{mcp-scan,mcp-scan-web}. Akto provides a platform to detect and protect against a set of attacks, such as prompt injection, tool poisoning, and data leaks. Palo Alto Networks integrates its Prisma AIRS platform into MCP to detect and defend against multiple Agent-specific threats and prompt injection categories across multiple languages~\cite{prisma-airs}. DataDome~\cite{datedome-mcp} provides a toolset to detect and prevent agent attacks with real-time visibility. MCPGuarian provides a security proxy to proxy to achieve real-time control over LLM interactions with MCP servers. Prompt Security provides a gateway to monitor and enforcement the MCP server developed in one organization~\cite{prompt-security}.

However, all these tools focus on the MCP server as the analysis and defend target, not the SDK implementation. Moreover, most of them rely on pre-defined attack templates collected by heuristics method to flag risks, which limits coverage and misses the potential risks of clause-backed omissions at the SDK boundary. 

In contrast, we propose a SDK and clause-centric analyzer that systematically enumerates omissions and assesses their exploitability via a modality-based analysis. Our tool fills the gap between application-level scanners and SDK/clause-level conformance. We are the first to validate clauses and code implementaions across SDKs, and surfaces root causes rather than template-based attack symptoms.

\section{Conclusion}
In this paper, we reveal a new attack surface intrinsic to the MCP's high-compatibility clause, motivated by the diversity feature of AI agents. These non-implemented clauses can be exploited to perform multiple attacks, posing significant security risks. To systematically analyze and exploit these clauses not implemented in MCP multi-language-SDKs, we propose an approach that universal the language-specific SDK into language-agnostic IR, and combines the static analysis method with large language model capabilities to identify those non-implemented clauses per SDK. Based on these, we perform an exploitable analysis to identify those potentially risky issues rather than normal bugs. Our approach has successfully identified 1,265 potential risks across ten SDKs. We have submitted 26 reports, 20 of which have been confirmed, including five high-priority risks. Furthermore, our tool has been invited for integration into an official MCP SEP. Together, these results highlight the practical impact of our work.

\section*{LLM Usage Considerations}
In this paper, we integrate the LLM into our approach pipeline. More specifically, we use LLM for reasoning and analyzing the non-implementation clauses across different MCP SDKs, and use an LLM-based pipeline for reasoning the exploitable non-implementation clauses. All usages follow the IEEE Secutiry\&Privacy LLM Policy. 

For originality, regarding the writing of this paper, LLMs are used only for editorial purposes in this manuscript. More specifically, we only use LLM to polish our writing for spelling or grammar issues. All technical ideas and related data are originally generated by the authors to ensure originality. Besides, by manually collecting related paper and writing literature review, not LLM, we are responsible for the thoroughness of their literature review and determining relevant prior work, and cite it to ensure proper credit. 

For transparency, we use LLMs' semantic understanding and reasoning capabilities to analyze and exploit those non-implemented clauses in MCP SDKs. In the non-implementation clause analyzer, the LLM only verifies whether the IR-level evidence indeed supports a “non-implementation” verdict. In the exploitability analysis, attack modalities are defined by the authors; the LLM only checks clause semantics vs. those modalities. In other words, we do not use LLM to generate new ideas. For result validation, we produce a three-expert validation for evaluating the non-implementation clause analyzer, and report the result of exploitability analysis to related communities and get multiple acknowledgments. All these validation methods make sure the LLM's result is validated. We also make a detailed discussion in the paper for the reason of using LLMs(section 4.2) and potential limitations(section 6.2). Our approach and results will also be open-sourced for reproducing and transparency. 

For responsibility, all data provided to the LLM are open-source MCP documents and code, and no personally identifiable information or copyrighted third-party text beyond what is permissible under scholarly fair use was used. From a security perspective, the LLM only checks semantic consistency between structured inputs (IR, source code, and clauses) and human-defined attack modalities, and every LLM judgment is accompanied by source-level evidence that can be reviewed by human experts. From a resource-usage and environmental perspective, we use existing LLMs only in inference mode and do not train any models. We design the pipeline to minimize the number and length of LLM queries (i.e., by guiding the interaction with IRs and enforcing a maximum number of LLM calls for bounded, auditable reasoning). All experiments are run on LLM Query APIs with limited costs (see more details in Section 5.1.4).



\bibliographystyle{IEEEtran}   
\bibliography{reference} 

\end{document}